# General Theory for the Ferroelectric Polarization Induced by Spin-Spiral Order

H. J. Xiang[1*], E. J. Kan[2], Y. Zhang[3], M.-H. Whangbo[3†], and X. G. Gong[1‡]

[1] Key Laboratory of Computational Physical Sciences (Ministry of Education), and Department of Physics, Fudan University, Shanghai 200433, P. R. China

[2] Department of Applied Physics, Nanjing University of Science and Technology, Nanjing, Jiangsu 210094, P. R. China

[3] Department of Chemistry, North Carolina State University, Raleigh, North Carolina 27695-8204, USA

**Abstract**

The ferroelectric polarization of triangular-lattice antiferromagnets induced by helical spin-spiral order is not explained by any existing model of magnetic-order-driven ferroelectricity. We resolve this problem by developing a general theory for the ferroelectric polarization induced by spin-spiral order and then by evaluating the coefficients needed to specify the general theory on the basis of density functional calculations. Our theory correctly describes the ferroelectricity of triangular-lattice antiferromagnets driven by helical spin-spiral order, and incorporates known models of magnetic-order-driven ferroelectricity as special cases.

Multiferroics, displaying magnetic, polar and elastic order parameters simultaneously, present fascinating fundamental physics [1,2] and potentially promising applications [3]. Spin-spiral multiferroics [1,4,5] constitute a challenging and interesting class of ferroelectricity in which the ferroelectric polarization **P** is induced by a magnetic order that removes inversion symmetry. For multiferroics with cycloidal spiral-spin order (e.g., TbMnO$_3$ [6-8] and MnWO$_4$ [9,10]), the ferroelectricity is explained by the inverse Dzyaloshinskii-Moriya (DM) interaction [11] or, equivalently, by the spin current model of Katsura, Nagaosa and Balatsky (KNB) [12], leading to **P**$_{ij}$ ∝ **e**$_{ij}$×(**S**$_i$×**S**$_j$), where **e**$_{ij}$ is a unit vector connecting the two adjacent spins **S**$_i$ and **S**$_j$. This model predicts that **P** is perpendicular to the direction of the magnetic modulation **q** ∝ **e**$_{ij}$ (i.e., **P** ⊥ **q**). Triangular-lattice antiferromagnets such as CuFeO$_2$ and AgCrO$_2$ also exhibit ferroelectricity when they adopt a helical spiral-spin order [13-15], in which the plane of the spin rotation is perpendicular to **q**. CuFeO$_2$ shows ferroelectric polarization when its magnetic structure has a helical spin-spiral order with **q** = (Q, Q, 0), where Q ≈ 1/3. The layered iodide MnI$_2$ was also found to be a multiferroic with helical spin-spiral order [16]. The experimental studies on CuFeO$_2$ and MnI$_2$ show that the **P** in the helical spin-spiral state with **q** = (Q, Q, 0) is parallel to **q** (i.e., **P** ∥ **q**). This finding is not explained either by the symmetric exchange striction mechanism or by the KNB model. The charge transfer between metal and ligand induced by spin-orbit coupling (SOC) was considered responsible for the ferroelectric polarization in a triangular lattice with helical spin-spiral order [17] with the prediction **P**$_{ij}$ ∝ (**e**$_{ij}$·**S**$_i$)**S**$_i$ − (**e**$_{ij}$·**S**$_j$)**S**$_j$. This polarization, known as the "bond polarization" [18], lies in the plane spanned by **S**$_i$ and **S**$_j$, which is perpendicular to **q**, and hence contradicts the experimental observation [14,16,19] that **P** ∥ **q** when **q** = (Q, Q, 0). In short, to explain the ferroelectric polarization of triangular-lattice antiferromagnets with helical spin-spiral order, it is necessary to develop a general theory for the ferroelectric polarization driven by spin-spiral order.

In this Letter we resolve the aforementioned issue first by developing a general theory for the ferroelectric polarization induced by spin-spiral on the basis of symmetry considerations and then by evaluating the coefficients needed to specify the general theory on the basis of density functional calculations for $MnI_2$ as a representative example. We demonstrate that our theory correctly describes the ferroelectric polarization of $MnI_2$, and the existing models of magnetic-order-driven ferroelectricity are special cases of our theory.

Let us first consider a spin dimer (i.e., a pair of adjacent spin sites) with spatial inversion symmetry at the center. Without loss of generality, the propagation vector from spin 1 to spin 2 will be taken along the x-axis. A noncollinear spin arrangement of the dimer removes the inversion symmetry and hence induces ferroelectric polarization **P**. In general, **P** is a function of the directions of spin 1 and spin 2 (with unit vectors **S**$_1$ and **S**$_2$, respectively), namely, $\mathbf{P} = \mathbf{P}(S_{1x}, S_{1y}, S_{1z}, S_{2x}, S_{2y}, S_{2z})$. In principle, therefore, **P** can be expanded as a Taylor series of $S_{i\alpha}$ (i = 1, 2; α = x, y, z). The time-reversal symmetry requires that inverting both spin directions leave the electric polarization unchanged. Thus, the odd terms of the Taylor expansion should vanish. If the fourth and higher order terms are neglected, **P** is written as

$$\mathbf{P} = \mathbf{P}_1(\mathbf{S}_1) + \mathbf{P}_2(\mathbf{S}_2) + \mathbf{P}_{12}(\mathbf{S}_1, \mathbf{S}_2), \qquad (1)$$

where the intra-site polarization $\mathbf{P}_i(\mathbf{S}_i)$ (i = 1, 2) and the inter-site polarization $\mathbf{P}_{12}(\mathbf{S}_1, \mathbf{S}_2)$ are given by

$$\mathbf{P}_i(\mathbf{S}_i) = \sum_{\alpha\beta} \mathbf{P}_i^{\alpha\beta} S_{i\alpha} S_{i\beta},$$

$$\mathbf{P}_{12}(\mathbf{S}_1, \mathbf{S}_2) = \sum_{\alpha\beta} \mathbf{P}_{12}^{\alpha\beta} S_{1\alpha} S_{2\beta}. \qquad (2)$$

Here the expansion coefficients, $\mathbf{P}_i^{\alpha\beta}$ and $\mathbf{P}_{12}^{\alpha\beta}$, are vectors. The above expressions show that $\mathbf{P}_i^{\alpha\beta} = \mathbf{P}_i^{\beta\alpha}$, $\mathbf{P}_i(\mathbf{S}_i) = \mathbf{P}_i(-\mathbf{S}_i)$, and $\mathbf{P}_{12}(-\mathbf{S}_1, \mathbf{S}_2) = \mathbf{P}_{12}(\mathbf{S}_1, -\mathbf{S}_2) = -\mathbf{P}_{12}(\mathbf{S}_1, \mathbf{S}_2)$. From these relationships, together with the use of spatial inversion symmetry and time-reversal symmetry, one can show that $\mathbf{P}_1^{\alpha\beta} = -\mathbf{P}_2^{\alpha\beta}$, and $\mathbf{P}_{12}^{\alpha\beta} = -\mathbf{P}_{12}^{\beta\alpha}$ [20]. The latter relation shows that the diagonal coefficients $\mathbf{P}_{12}^{\alpha\alpha} = 0$, so the inter-site polarization can be expressed as

$$\mathbf{P}_{12} = \mathbf{P}_{12}^{yz}\,(\mathbf{S}_1 \times \mathbf{S}_2)_x + \mathbf{P}_{12}^{zx}\,(\mathbf{S}_1 \times \mathbf{S}_2)_y + \mathbf{P}_{12}^{xy}\,(\mathbf{S}_1 \times \mathbf{S}_2)_z, \tag{3a}$$

where $(\mathbf{S}_1 \times \mathbf{S}_2)_\alpha$ refers to the $\alpha$ (= x, y, z) component of the vector $(\mathbf{S}_1 \times \mathbf{S}_2)$. Using similar notations for the x, y and z components of the vectors $\mathbf{P}_{12}^{\alpha\beta}$, Eq. 3a is rewritten as

$$\mathbf{P}_{12} = \mathbf{M}\,(\mathbf{S}_1 \times \mathbf{S}_2) \tag{3b}$$

using the 3×3 matrix $\mathbf{M}$

$$\mathbf{M} = \begin{bmatrix} (\mathbf{P}_{12}^{yz})_x & (\mathbf{P}_{12}^{zx})_x & (\mathbf{P}_{12}^{xy})_x \\ (\mathbf{P}_{12}^{yz})_y & (\mathbf{P}_{12}^{zx})_y & (\mathbf{P}_{12}^{xy})_y \\ (\mathbf{P}_{12}^{yz})_z & (\mathbf{P}_{12}^{zx})_z & (\mathbf{P}_{12}^{xy})_z \end{bmatrix}. \tag{4}$$

Given that the propagation vector from spin 1 to spin 2 is taken along the x-axis, the bond polarization model [18] is a special case of the intra-site polarization in which the only nonzero coefficients are $\mathbf{P}_1^{xx} = (C, 0, 0)$, $\mathbf{P}_1^{xy} = \mathbf{P}_1^{yx} = (0, C/2, 0)$ and $\mathbf{P}_1^{zx} = \mathbf{P}_1^{xz} = (0, 0, C/2)$, where C is a constant. The KNB model is a special case of the inter-site polarization with $(\mathbf{P}_{12}^{zx})_z = -(\mathbf{P}_{12}^{xy})_y = C$ as the only nonzero elements of $\mathbf{M}$, where C is a constant. The inter-site polarization given by Eq. 3b may now be referred to as the generalized KNB (gKNB) model. For a linear three-atom M-L-M model (M = transition-metal, L = main-group ligand), the intra-site term reduces to the bond polarization model, and the inter-site term to the KNB model.

To specify the intra-site and inter-site polarizations described above, one needs to determine the expansion coefficients $\mathbf{P}_i^{\alpha\beta}$ (i = 1, 2) and $\mathbf{P}_{12}^{\alpha\beta}$. We evaluate these coefficients for a spin dimer of $MnI_2$ (Fig. 1(a)) as a representative example, on the basis of density functional calculations. We adopt the LDA+U+SOC approach to calculate electric polarizations [20]. $MnI_2$ crystallizes in the $CdI_2$ type structure with $MnI_2$ layer stacked along the c axis [see the left inset of Fig. 1(a)]. In the Mn triangular lattice, each $Mn^{2+}$ ion has six nearest neighbor (NN) $Mn^{2+}$ ions. The structure of an isolated $Mn_2I_{10}$ cluster (i.e. a spin dimer), namely, an isolated Mn-Mn pair plus its 10 first-coordinate I atoms, is shown in the upper-right inset of Fig. 1(a). Each NN Mn-Mn pair contributes to the total electric polarization. To characterize the ferroelectric polarization arising from one pair of NN $Mn^{2+}$ ions in $MnI_2$, we isolate a Mn-Mn pair in a 5×5×1 supercell of $MnI_2$ and replace all other $Mn^{2+}$ ions with nonmagnetic $Mg^{2+}$ ions, as depicted in Fig. 1(a). (A more accurate method for calculating the coefficients of the inter-site term requires no substitution of $Mn^{2+}$ ions with nonmagnetic ions such as $Mg^{2+}$ ions [20], and will be referred to as the no-substitution method.) When the SOC effect is excluded in the density functional calculations, the electric polarizations become zero so that the SOC effect is essential for the occurrence of ferroelectricity in helical spin-spiral systems.

The expansion coefficients $\mathbf{P}_i^{\alpha\beta}$ (i = 1, 2) and $\mathbf{P}_{12}^{\alpha\beta}$ for a given spin dimer can be readily determined by mapping analysis once its polarizations are calculated for a set of carefully-chosen noncollinear spin arrangements. To evaluate an off-diagonal coefficient of the intra-site polarization, for example, $\mathbf{P}_1^{xy}$, we calculate the electric polarizations for the four spin arrangements I′ − IV′ of the spin dimer specified in Table I. Then, according to Eq. 2, $\mathbf{P}_1^{xy}$ is related to the polarization of the spin arrangements I′ − IV′ as $\mathbf{P}_1^{xy} = (\mathbf{P}_{I'} + \mathbf{P}_{II'} − \mathbf{P}_{III'} − \mathbf{P}_{IV'})/4$.

Other off-diagonal intra-site coefficients, $\mathbf{P}_1^{xz}$ and $\mathbf{P}_1^{yz}$, can be evaluated in a similar manner. The diagonal coefficients of the intra-site polarization can be determined by calculating the electric polarizations for the six spin arrangements I – VI of the spin dimer specified in Table II. According to Eq. 2, the polarizations of these spin arrangements have the relationships, $\mathbf{P}_I + \mathbf{P}_{II} = 2(\mathbf{P}_1^{xx} - \mathbf{P}_1^{yy})$, $\mathbf{P}_{III} + \mathbf{P}_{IV} = 2(\mathbf{P}_1^{xx} - \mathbf{P}_1^{zz})$, and $\mathbf{P}_V + \mathbf{P}_{VI} = 2(\mathbf{P}_1^{yy} - \mathbf{P}_1^{zz})$. Two of these three equations are linearly independent, but only the two independent parameters $(\mathbf{P}_1^{xx} - \mathbf{P}_1^{yy})$ and $(\mathbf{P}_1^{xx} - \mathbf{P}_1^{zz})$ are needed in calculating the sum of the diagonal contributions of the two intra-site polarizations because of the relationship $\mathbf{P}_1^{\alpha\beta} = -\mathbf{P}_2^{\alpha\beta}$ [20]. The electric polarizations of the above six spin arrangements can also be used to extract the coefficients of the inter-site polarization $\mathbf{P}_{12}$, that is, $\mathbf{P}_{12}^{xy} = (\mathbf{P}_I - \mathbf{P}_{II})/2$, $\mathbf{P}_{12}^{xz} = (\mathbf{P}_{III} - \mathbf{P}_{IV})/2$, $\mathbf{P}_{12}^{yz} = (\mathbf{P}_V - \mathbf{P}_{VI})/2$.

Our calculations for the spin dimer of $MnI_2$ and mapping analyses as outlined above show that the coefficients of the intra-site polarization are $\mathbf{P}_1^{xx} = (0,0,0)$, $\mathbf{P}_1^{yy} = (2.5, 0, 0)$, $\mathbf{P}_1^{zz} = (-2.5, 0, 0)$, $\mathbf{P}_1^{xy} = (5.0, 7.5, 0)$, $\mathbf{P}_1^{xz} = (0, -5.0, 0)$, and $\mathbf{P}_1^{yz} = (7.5, -2.5, 0)$ in units of $10^{-6}$ eÅ. Note that the expression of the intra-site polarization differs from that of the bond polarization model (see above). The coefficients of the inter-site polarization extracted by using the no-substitution method [20] are

$$\mathbf{M} = \begin{bmatrix} M_{11} & 0 & 0 \\ 0 & M_{22} & M_{23} \\ 0 & M_{32} & M_{33} \end{bmatrix}, \quad (5)$$

where, in units of $10^{-5}$ eÅ, $M_{11} = -4.8$, $M_{22} = 39.5$, $M_{23} = 49.0$, $M_{32} = -44.5$, and $M_{33} = -26.0$. Thus, the inter-site polarization is at least an order-of-magnitude stronger than the intra-site polarization, and differs from the KNB model (see above) because the matrix elements $M_{11} =$

$(\mathbf{P}_{12}^{yz})_x$, $M_{22} = (\mathbf{P}_{12}^{zx})_y$, and $M_{33} = (\mathbf{P}_{12}^{xy})_z$ are not zero and because $M_{23} = (\mathbf{P}_{12}^{xy})_y$ is different from $-M_{32} = -(\mathbf{P}_{12}^{zx})_z$. Fig. 1(b) illustrates the differences between the KNB and gKNB models in predicting the polarization **P** for three different spin arrangements of the Mn-Mn dimer. Given the expansion coefficients $\mathbf{P}_{12}^{\alpha\beta}$ extracted as described above, one can predict the **P** of the Mn-Mn dimer with various spin arrangements by using Eq. 1. To show that the gKNB model can indeed predict the **P** of the Mn-Mn pair with arbitrary spin orientations, we compute **P** for several spin arrangements of the Mn-Mn dimer directly from density functional calculations. For convenience, we keep the first Mn spin along the x direction and rotate the second Mn spin in the xy-plane in these spin arrangements. Then, the **P** is found to lie in the yz-plane. Importantly, the polarization predicted by the gKNB model is in excellent agreement with the value calculated directly from density functional calculations for the spin dimer [see Fig. 1(c)]. This validates our analysis of the electric polarization without considering the fourth and higher order terms.

With the electric polarizations calculated for various NN Mn-Mn pairs, we now estimate the electric polarization of $MnI_2$ with helical spin-spiral order in terms of only the inter-site term, because the sum of all intra-site terms for any helical spin-spiral arrangement is zero. Since each Mn spin site i has six NN Mn spins k (= 1 – 6), the total polarization $\mathbf{P}_i^{tot}$ at the site i is written as $\mathbf{P}_i^{tot} = \sum_{k=1}^{6} \mathbf{P}_{ik}$. In the case of spin-spiral, $\mathbf{P}_i^{tot}$ is the same for all i sites, so we consider only the polarization associated with site 0 shown in Fig. 2(a), for which $\mathbf{P}_i^{tot} = \sum_{k=1}^{6} \mathbf{P}_{0k} = \sum_{k=1}^{6} \mathbf{M}^{0k}(\mathbf{S}_0 \times \mathbf{S}_k)$, where $\mathbf{M}^{0k}$ refers to the matrix for the inter-site polarization for pair 0 and k. In the local (x, y, z) coordinate system defined in Fig. 1(a), our calculations show that for $\mathbf{q} = (Q, 0, 0)$, $\mathbf{P}_0^{tot} = (\frac{\sqrt{3}}{2}A, -\frac{3}{2}A, 0)$ with $A = (M_{11} - M_{22}) \sin 2\pi Q$. In the case of $\mathbf{q} = (Q,$

Q, 0), $\mathbf{P}_0^{tot}$ = ($\frac{1}{2}$B, $\frac{\sqrt{3}}{2}$B, 0) with B = ($M_{11}$ + 3 $M_{22}$ − 4 $M_{11}$ cos2πQ) sin2πQ. Thus, the gKNB model predicts that $\mathbf{P} \perp \mathbf{q}$ when $\mathbf{q}$ = (Q, 0, 0), but $\mathbf{P} \parallel \mathbf{q}$ in the case of $\mathbf{q}$ = (Q, Q, 0), as found experimentally [16], and that the polarization reverses with the change in the spin chirality ($\mathbf{q}$ to −$\mathbf{q}$), in accord with experiment. The gKNB model shows that the polarization in both cases depends only on two elements of the matrix $\mathbf{M}$, i.e., $M_{11}$ and $M_{22}$, both of which are zero in the KNB model. In Fig. 2(b), we plot the magnitude of the polarization as a function of Q for the cases of $\mathbf{q}$ = (Q, 0, 0) and $\mathbf{q}$ = (Q, Q, 0). The plot is symmetric with maximum at Q = 0.25 in the case of $\mathbf{q}$ = (Q, 0, 0), but is slightly asymmetric with maximum at Q = 0.225 in the case of $\mathbf{q}$ = (Q, Q, 0).

We determine the total ferroelectric polarization of $MnI_2$ in the helical spin-spiral state with $\mathbf{q}$ = (0.181, 0, 0.439), observed in the absence of applied magnetic field, directly from density functional calculations by approximating the incommensurate state with the commensurate helical spin-spiral state with $\mathbf{q}$ = (1/3, 0, 0) using a 3×1×1 supercell. Our calculations show that the electric polarization of this state is 58.8 μC/m$^2$ along the [100] direction, as shown in Fig. 2(a). Thus, our density functional calculations show that $\mathbf{P} \perp \mathbf{q}$, in agreement with experiment [16]. For the helical spin-spiral state of $MnI_2$ with $\mathbf{q}$ = (Q, Q, 0), found under in-plane magnetic field greater than 3 T [16], we use a $\sqrt{3}\times\sqrt{3}\times 1$ supercell to simulate the $\mathbf{q}$ = (1/3, 1/3, 0) state. The total polarization of this state is calculated to be 71.4 μC/m$^2$ along the [110] direction. In this case, $\mathbf{P} \parallel \mathbf{q}$, again in agreement with experiment [16]. As can be seen from Figs. 2(c) and (d), the gKNB model not only predicts the correct direction of the polarization, but also gives a rather accurate magnitude of the polarization for the cases of $\mathbf{q}$ = (Q, 0, 0) and $\mathbf{q}$ = (Q, Q, 0). Our theory of ferroelectric polarization is general and is expected to provide accurate predictions when applied to other multiferroics driven by spin-spiral magnetic order.

In the local coordinate system (X, Y, Z) chosen to minimize the magnitudes of the diagonal elements of the matrix **M** [see the lower-right inset of Fig. 1(a), the Y axis is close to the distance vector between the two I atoms forming the shared octahedral edge between the adjacent Mn atoms], the matrix **M** of Eq. 5 determined from density functional calculations is rewritten as

$$\mathbf{M} = \begin{bmatrix} -4.8 & 0 & 0 \\ 0 & 6.8 & 79.6 \\ 0 & -13.9 & 6.8 \end{bmatrix}. \tag{6}$$

in units of $10^{-5}$ eÅ. In the local (X, Y, Z) coordinate system, $(\mathbf{P}_{12}^{XY})_Y = 79.6 \times 10^{-5}$ eÅ is much greater than $-(\mathbf{P}_{12}^{ZX})_Z = 13.9 \times 10^{-5}$ eÅ. The cause for this anisotropy was examined by performing tight-binding calculations for a planar $M_2L_2$ cluster consisting of two transition metal atoms M bridged by two ligand atoms L [20] on the basis of the model Hamiltonian similar to that employed by Jia *et al*. [18]. This analysis shows [20] that the large difference between $(\mathbf{P}_{12}^{XY})_Y$ and $-(\mathbf{P}_{12}^{ZX})_Z$ arises from the structural anisotropy of the planar $M_2L_2$ cluster; the Y axis is nearly in the plane of, but the Z axis is nearly perpendicular to, the plane of the cluster.

In summary, on the basis of symmetry arguments, we developed a general theory of ferroelectric polarization that can correctly describe all known ferroelectric polarization induced by spin-spiral order.

Work at Fudan was partially supported by NSFC, Pujiang plan, and Program for Professor of Special Appointment (Eastern Scholar).

* Electronic address: hxiang@fudan.edu.cn

† Electronic address: mike_whangbo@ncsu.edu

‡ Electronic address: xggong@fudan.edu.cn

Table I. The four spin arrangements I′ – IV′ of the spin dimer employed to calculate its off-diagonal intra-site electric polarization $\mathbf{P}_1^{xy}$ by LDA+U+SOC calculations.

|      | $\mathbf{S}_1$ | $\mathbf{S}_2$ |
|------|----------------|----------------|
| I′   | $(\frac{\sqrt{2}}{2}, \frac{\sqrt{2}}{2}, 0)$ | (1, 0, 0) |
| II′  | $(\frac{\sqrt{2}}{2}, \frac{\sqrt{2}}{2}, 0)$ | (−1, 0, 0) |
| III′ | $(\frac{\sqrt{2}}{2}, -\frac{\sqrt{2}}{2}, 0)$ | (1, 0, 0) |
| IV′  | $(\frac{\sqrt{2}}{2}, -\frac{\sqrt{2}}{2}, 0)$ | (−1, 0, 0) |

Table II. The six spin arrangements I – VI of the spin dimer employed to calculate its diagonal intra-site electric polarization $\mathbf{P}_1^{\alpha\alpha}$ ($\alpha$ = x, y, z) as well as the inter-site polarization $\mathbf{P}_{12}^{xy}$, $\mathbf{P}_{12}^{xz}$ and $\mathbf{P}_{12}^{yz}$ by LDA+U+SOC calculations.

|      | $\mathbf{S}_1$ | $\mathbf{S}_2$ |
|------|----------------|----------------|
| I    | (1, 0, 0)      | (0, 1, 0)      |
| II   | (1, 0, 0)      | (0, −1, 0)     |
| III  | (1, 0, 0)      | (0, 0, 1)      |
| IV   | (1, 0, 0)      | (0, 0, −1)     |
| V    | (0, 1, 0)      | (0, 0, 1)      |
| VI   | (0, 1, 0)      | (0, 0, −1)     |

**Figure captions**

Figure 1.   (Color online) (a) The 5×5×1 supercell of MnI$_2$ in which all Mn$^{2+}$ ions except for an isolated NN Mn-Mn pair are replaced by nonmagnetic Mg$^{2+}$ ions. The left inset illustrates the layered structure of MnI$_2$. The upper-right inset shows the top view of the Mn$_2$I$_{10}$ dimer cluster. The lower-right inset shows the local coordinate systems (x, y, z) and (X, Y, Z) employed for calculations. (b) The electric polarizations predicted by the KNB and gKNB models for three different spin configurations of the Mn-Mn dimer, where the directions of the spins and the polarizations are described in terms of the (x, y, z) coordinate system shown in Fig. 1(a). The blue dots representing **S**$_2$ means that it is pointed along the positive z-axis, and so does the green dot representing the polarization in the KNB model. The Cartesian components of the polarizations obtained from the gKNB model are given in units of 10$^{-5}$ eÅ. (c) The polarization of the Mn-Mn pair with spins in the xy-plane as a function of the angle α between the spins **S**$_1$ and **S**$_2$. The data points were obtained from direct density functional calculations, and the solid curves from the model of Eq. 1.

Figure 2.   (Color online) (a) The triangular lattice of Mn$^{2+}$ ions, where the in-plane lattice vectors a$_1$ and a$_2$ and the corresponding reciprocal lattice vectors b$_1$ and b$_2$ are shown. (b) The magnitude of the polarization predicted from our gKNB model as a function of Q for the cases of **q** = (Q, 0, 0) and **q** = (Q, Q, 0). (c, d) The spin orientations of two proper-screw spirals with **q** = (1/3, 0, 0) and **q** = (1/3, 1/3, 0). The modulation vector **q** and the polarization vector **P** are represented by the white and green arrows, respectively. The numbers (in μC /m$^2$) denote the magnitudes of the polarizations obtained from the direct density functional calculation and the gKNB model.

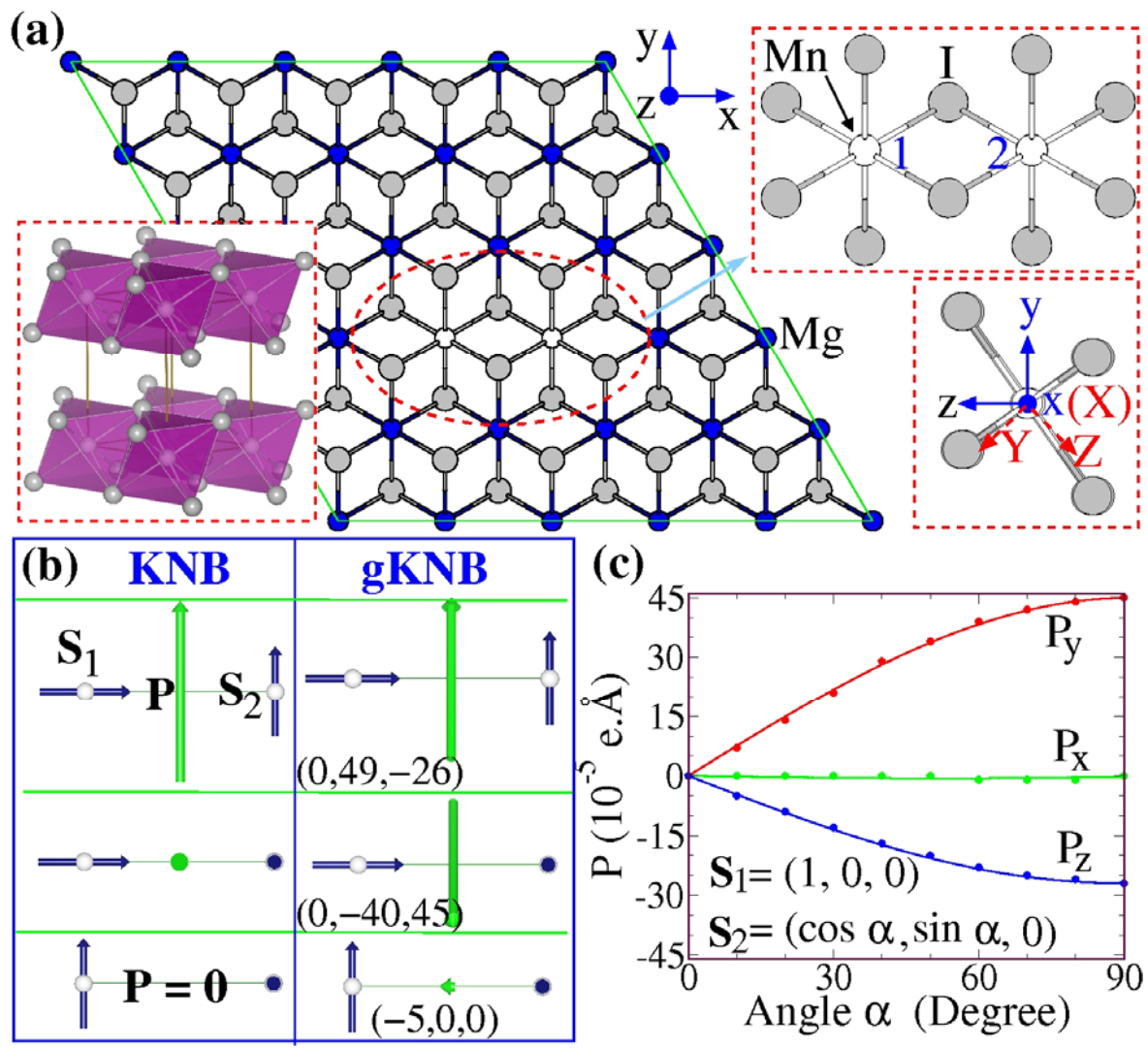

Figure 1.

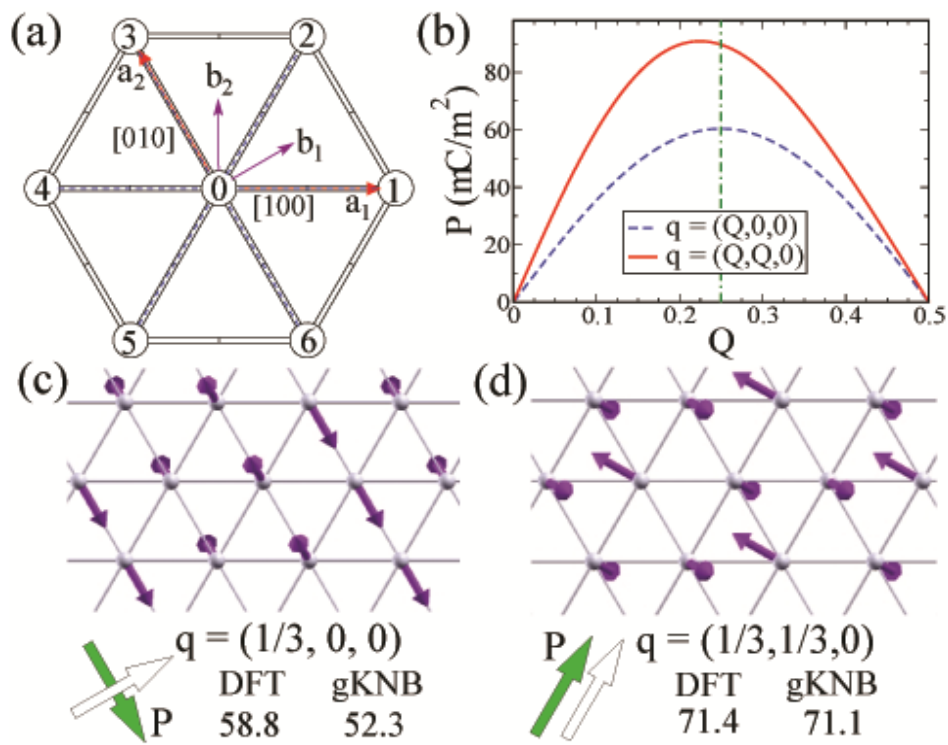

Figure 2

Supplementary Materials for

**General Theory for the Ferroelectric Polarization Induced by Spin-Spiral Order**

H. J. Xiang, E. J. Kan, Y. M. Zhang, M.-H. Whangbo, and X. G. Gong

## 1. Relationship between the coefficients of the electric polarization model

The ferroelectric polarization **P** of a spin dimer is written as

$$\mathbf{P} = \mathbf{P}_1(\mathbf{S}_1) + \mathbf{P}_2(\mathbf{S}_2) + \mathbf{P}_{12}(\mathbf{S}_1, \mathbf{S}_2), \tag{1}$$

where the intra-site polarization $\mathbf{P}_i(\mathbf{S}_i)$ (i = 1, 2) and the inter-site polarization $\mathbf{P}_{12}(\mathbf{S}_1, \mathbf{S}_2)$ are given by

$$\mathbf{P}_i(\mathbf{S}_i) = \sum_{\alpha\beta} \mathbf{P}_i^{\alpha\beta} S_{i\alpha} S_{i\beta},$$

$$\mathbf{P}_{12}(\mathbf{S}_1, \mathbf{S}_2) = \sum_{\alpha\beta} \mathbf{P}_{12}^{\alpha\beta} S_{1\alpha} S_{2\beta}, \tag{2}$$

where the expansion coefficients, $\mathbf{P}_i^{\alpha\beta}$ and $\mathbf{P}_{12}^{\alpha\beta}$, are vectors. The above expressions show that

$$\mathbf{P}_i^{\alpha\beta} = \mathbf{P}_i^{\beta\alpha},$$

$$\mathbf{P}_i(\mathbf{S}_i) = \mathbf{P}_i(-\mathbf{S}_i),$$

$$\mathbf{P}_{12}(-\mathbf{S}_1, \mathbf{S}_2) = \mathbf{P}_{12}(\mathbf{S}_1, -\mathbf{S}_2) = -\mathbf{P}_{12}(\mathbf{S}_1, \mathbf{S}_2). \tag{3}$$

To prove that show that $\mathbf{P}_1^{\alpha\beta} = -\mathbf{P}_2^{\alpha\beta}$, we consider the two spin arrangements of the spin dimer.

Arrangement I:   $\mathbf{S}_1 = \mathbf{S}_2 = \mathbf{S} = (S_x, S_y, S_z)$

Arrangement II:   $\mathbf{S}_1 = -\mathbf{S}_2 = \mathbf{S} = (S_x, S_y, S_z)$

The electric polarizations of both arrangements are zero because of the spatial inversion symmetry and time-reversal symmetry. These two configurations have opposite contribution to the inter-site polarization $\mathbf{P}_{12}$. Therefore, the sum $\mathbf{P}_{sum}$ of the electric polarizations of these two spin arrangements only contains the intra-site polarizations.

$$\mathbf{P}_{sum} = 2\mathbf{P}_1(\mathbf{S}) + 2\mathbf{P}_2(\mathbf{S}) = 2\sum_{\alpha\beta} (\mathbf{P}_1^{\alpha\beta} + \mathbf{P}_2^{\alpha\beta}) S_\alpha S_\beta. \tag{4}$$

Since both spin arrangements have zero electric polarization, $\mathbf{P}_{sum}$ should be zero. Because the spin direction **S** is arbitrary, we obtain

$$\mathbf{P}_1^{\alpha\beta} = -\mathbf{P}_2^{\alpha\beta}. \tag{5}$$

To prove that $\mathbf{P}_{12}^{\alpha\beta} = -\mathbf{P}_{12}^{\beta\alpha}$, we consider two spin arrangements.

Arrangement I:       $\mathbf{S}_1 = \mathbf{S}$ and $\mathbf{S}_2 = \mathbf{S}'$

Arrangement II:     $\mathbf{S}_1 = \mathbf{S}'$ and $\mathbf{S}_2 = \mathbf{S}$.

It is noted that arrangement II is obtained by performing the spatial inversion operation on arrangement I. Thus, arrangement II has an electric polarization opposite to that of arrangement I. The inter-site polarization of arrangement II should be also opposite to that of arrangement I, namely,

$$\sum_{\alpha\beta} \mathbf{P}_{12}^{\alpha\beta} S_\alpha S'_\beta = -\sum_{\alpha\beta} \mathbf{P}_{12}^{\alpha\beta} S'_\alpha S_\beta. \tag{6}$$

The above equation shows that $\mathbf{P}_{12}^{\alpha\beta} = -\mathbf{P}_{12}^{\beta\alpha}$, and hence $\mathbf{P}_{12}^{\alpha\alpha} = 0$.

## 2. Details of the density functional calculations

Total energy calculations are based on the DFT plus the on-site repulsion (U) method [1] within the local density approximation (LDA+U) on the basis of the projector augmented wave method [2] encoded in the Vienna ab initio simulation package [3]. The plane-wave cutoff energy is set to 400 eV. Spin-orbit coupling (SOC) is included in the calculations unless noted otherwise. We mainly discuss the results obtained with the on-site repulsion $U = 5$ eV and the exchange parameter $J = 1$ eV on Mn. We also did LDA+SOC calculations to find that the main results are qualitatively similar. For the calculation of electric polarization, the Berry phase method [4] was used.

## 3. Sum of the diagonal contributions of the two intra-site polarizations

The sum $\Sigma$ of the diagonal contributions of the two intra-site polarizations is expressed as

$$\Sigma = \mathbf{P}_1^{xx} S_{1x} S_{1x} + \mathbf{P}_1^{yy} S_{1y} S_{1y} + \mathbf{P}_1^{zz} S_{1z} S_{1z} + \mathbf{P}_2^{xx} S_{2x} S_{2x} + \mathbf{P}_2^{yy} S_{2y} S_{2y} + \mathbf{P}_2^{zz} S_{2z} S_{2z} \qquad (1)$$

Because of the relationship

$$\mathbf{P}_1^{\alpha\beta} = -\mathbf{P}_2^{\alpha\beta}, \qquad (2)$$

Eq. 1 can be rewritten as

$$\Sigma = (\mathbf{P}_1^{xx} - \mathbf{P}_1^{zz}) S_{1x} S_{1x} + (\mathbf{P}_1^{yy} - \mathbf{P}_1^{zz}) S_{1y} S_{1y} - (\mathbf{P}_1^{xx} - \mathbf{P}_1^{zz}) S_{2x} S_{2x} - (\mathbf{P}_1^{yy} - \mathbf{P}_1^{zz}) S_{2y} S_{2y}. \qquad (3)$$

Consequently, only the two independent parameters ($\mathbf{P}_1^{xx} - \mathbf{P}_1^{yy}$) and ($\mathbf{P}_1^{xx} - \mathbf{P}_1^{zz}$) are needed in calculating the sum $\Sigma$.

**4. Calculations of the electric polarizations of a spin dimer**

**4.1. Using the substitution method**

As described in the text, this substitution method uses a 5×5×1 supercell of $MnI_2$ to define a spin dimer, namely, the 23 of the 25 $Mn^{2+}$ ions in the supercell (except for the adjacent two defining a spin dimer) are replaced with nonmagnetic $Mg^{2+}$ ions.

The electric polarizations (in unit of eÅ) calculated for the spin arrangements I′ − IV′ of the spin dimer (see Table I) were calculated to extract the off-diagonal term $\mathbf{P}_1^{xy} = (\mathbf{P}_{I'} + \mathbf{P}_{II'} - \mathbf{P}_{III'} - \mathbf{P}_{IV'})/4$:

    I′:    (0.00001000, -0.00031000, 0.00019000)

    II′:   (0.00001000, 0.00033000, -0.00019000)

    III′:  (0.00000000, 0.00032000, -0.00019000)

    IV′:  (0.00000000, -0.00033000, 0.00019000)

Similarly, the electric polarizations were calculated for the four spin arrangements shown below to extract off-diagonal term $\mathbf{P}_1^{xz} = (\mathbf{P}_{I'} + \mathbf{P}_{II'} - \mathbf{P}_{III'} - \mathbf{P}_{IV'})/4$:

|      | $S_1$ | $S_2$ |
|------|-------|-------|
| I′   | $(\frac{\sqrt{2}}{2}, 0, \frac{\sqrt{2}}{2})$ | (1, 0, 0) |
| II′  | $(\frac{\sqrt{2}}{2}, 0, \frac{\sqrt{2}}{2})$ | (−1, 0, 0) |
| III′ | $(\frac{\sqrt{2}}{2}, 0, -\frac{\sqrt{2}}{2})$ | (1, 0, 0) |
| IV′  | $(\frac{\sqrt{2}}{2}, 0, -\frac{\sqrt{2}}{2})$ | (−1, 0, 0) |

I':   (0.00000000, 0.00025000, -0.00034000)

II':  (0.00000000, -0.00026000, 0.00034000)

III': (0.00000000, -0.00025000, 0.00034000)

IV':  (0.00000000, 0.00026000, -0.00034000)

The electric polarizations were calculated for the four spin arrangements shown below to extract off-diagonal term $\mathbf{P}_1^{yz} = (\mathbf{P}_{I'} + \mathbf{P}_{II'} - \mathbf{P}_{III'} - \mathbf{P}_{IV'})/4$:

|      | $\mathbf{S}_1$ | $\mathbf{S}_2$ |
|------|----------------|----------------|
| I'   | $(0, \frac{\sqrt{2}}{2}, \frac{\sqrt{2}}{2})$ | $(1, 0, 0)$ |
| II'  | $(0, \frac{\sqrt{2}}{2}, \frac{\sqrt{2}}{2})$ | $(-1, 0, 0)$ |
| III' | $(0, \frac{\sqrt{2}}{2}, -\frac{\sqrt{2}}{2})$ | $(1, 0, 0)$ |
| IV'  | $(0, \frac{\sqrt{2}}{2}, -\frac{\sqrt{2}}{2})$ | $(-1, 0, 0)$ |

I':   (0.00001000, -0.00007000, -0.00015000)

II':  (0.00001000, 0.00007000, 0.00015000)

III': (0.00000000, -0.00057000, 0.00051000)

IV':  (-0.00001000, 0.00058000, -0.00051000)

The electric polarizations were calculated for the six spin arrangements I – VI (see Table II) to extract the diagonal intra-site terms $\mathbf{P}_1^{xx}$, $\mathbf{P}_1^{yy}$ and $\mathbf{P}_1^{zz}$ as well as the inter-site terms $\mathbf{P}_{12}^{xy}$, $\mathbf{P}_{12}^{xz}$ and $\mathbf{P}_{12}^{yz}$:

| | |
|---|---|
| I: | (0, 0.00045, -0.00027) |
| II: | (0, -0.00045, 0.00026) |
| III: | (0, -0.00036, 0.00047) |
| IV: | (0, 0.00036, -0.00047) |
| V: | (-0.00004, 0, 0) |
| VI: | (0.00005, 0, 0) |

In units of eÅ, the above results give rise to $M_{11}$ = -0.000045, $M_{22}$ = 0.00036, $M_{23}$ = 0.00045, $M_{32}$ = -0.00047 and $M_{33}$ = -0.000265.

### 4.2. Using the no-substitution method

The substitution method has an undesirable effect in that it distorts the electron distribution around the spin dimer, because the valence atomic orbitals of Mn differ from those of Mg. To evaluate the inter-site polarizations more accurately, therefore, we employ the "magnetic" spin-dimer method, which is similar to the technique that we recently proposed to extract the spin exchange parameters (see: H. J. Xiang, E. J. Kan, S.-H. Wei, M.-H. Whangbo and X. G. Gong, arXiv:1106.5549).

To obtain the inter-site polarization $\mathbf{P}_{12}^{xy}$, for example, we use a 5×5×1 supercell of $MnI_2$ with 25 $Mn^{2+}$ ions. The first two $Mn^{2+}$ ions (1 and 2) will be regarded as the spin dimer for which the inter-site polarizations are to be extracted. We calculate the electric polarizations of the following four spin arrangements A – D of the supercell:

|   | $S_1$ | $S_2$ | 23 other spins |
|---|-------|-------|----------------|
| A | (1, 0, 0)  | (0, 1, 0)  | (0, 0, 1) |
| B | (1, 0, 0)  | (0, -1, 0) | (0, 0, 1) |
| C | (-1, 0, 0) | (0, 1, 0)  | (0, 0, 1) |
| D | (-1, 0, 0) | (0, -1, 0) | (0, 0, 1) |

By using Eq. 2 of the text, it can be easily shown that $\mathbf{P}_{12}^{xy} = (\mathbf{P}_A + \mathbf{P}_D - \mathbf{P}_B - \mathbf{P}_C)/4$. The inter-site polarizations for all other Mn-Mn pairs cancel out, and so do all intra-site polarizations. The calculated polarizations for the four spin arrangements are:

$\mathbf{P}_A = (-0.000040, \ 0.000880, \ -0.000720)$

$\mathbf{P}_B = (\ 0.000040, \ -0.000100, \ -0.000200)$

$\mathbf{P}_C = (-0.000040, \ -0.000880, \ \ 0.000720)$

$\mathbf{P}_D = (\ 0.000040, \ \ 0.000100, \ \ 0.000200)$

so that, in units of $10^{-5}$ eÅ, we obtain $\mathbf{P}_{12}^{xy} = (\mathbf{P}_A + \mathbf{P}_D - \mathbf{P}_B - \mathbf{P}_C)/4 = (0, 49.0, -26.0)$.

To extract $\mathbf{P}_{12}^{yz}$, the polarizations were calculated for the four spin arrangements:

|   | $S_1$ | $S_2$ | 23 other spins |
|---|-------|-------|----------------|
| A | (0, 1, 0)  | (0, 0, 1)  | (0, 0, 1) |
| B | (0, 1, 0)  | (0, 0, -1) | (0, 0, 1) |
| C | (0, -1, 0) | (0, 0, 1)  | (0, 0, 1) |
| D | (0, -1, 0) | (0, 0, -1) | (0, 0, 1) |

$\mathbf{P}_A = (0, 0, 0)$

$\mathbf{P}_B = (0.000070, \ 0.000000, \ 0.000000)$

$\mathbf{P}_C = (0, 0, 0)$

$\mathbf{P}_D = (-0.000120, \ 0.000000, \ 0.000000)$

In units of $10^{-5}$ eÅ, these values lead to $\mathbf{P}_{12}^{yz} = (-4.8, \ 0, \ 0)$.

To extract $\mathbf{P}_{12}^{xz}$, the polarizations were calculated for the four spin arrangements:

|   | $\mathbf{S}_1$ | $\mathbf{S}_2$ | 23 other spins |
|---|---|---|---|
| A | (1, 0, 0) | (0, 0, 1) | (0, 0, 1) |
| B | (1, 0, 0) | (0, 0, -1) | (0, 0, 1) |
| C | (-1, 0, 0) | (0, 0, 1) | (0, 0, 1) |
| D | (-1, 0, 0) | (0, 0, -1) | (0, 0, 1) |

$\mathbf{P}_A = (0, 0, 0)$

$\mathbf{P}_B = (-0.000010, \ 0.000790, \ -0.000890)$

$\mathbf{P}_C = (0, 0, 0)$

$\mathbf{P}_D = (-0.000020, \ -0.000790, \ 0.000890)$.

In units of $10^{-5}$ eÅ, we get $\mathbf{P}_{12}^{xz} = (\mathbf{P}_A + \mathbf{P}_D - \mathbf{P}_B - \mathbf{P}_C)/4 = (-0.3, \ -39.5, \ 44.5)$. Note that $\mathbf{P}_{12}^{xz} = -\mathbf{P}_{12}^{zx}$. Therefore, in units of $10^{-5}$ eÅ, $M_{11} = -4.8$, $M_{22} = 39.5$, $M_{23} = 49.0$, $M_{32} = -44.5$, and $M_{33} = -26.0$.

## 5. Details of the tight-binding calculations

In our tight-binding (TB) model, we consider two transition metal atoms ($M_l$ and $M_r$) bridged by two ligand anions ($L_u$ and $L_d$), as shown in Fig. S1(a). Our model is similar to that proposed by Jia et al.[1] except that they considered a linear M-L-M three-atom model. The overall Hamiltonian describing the four-atom cluster is given by

$$H = H_M + H_L + H_t + H_{so}, \tag{S1}$$

where

$$H_M = \sum_a^{l,r} \sum_\sigma \left( \sum_{\alpha \in t_{2g}} E_{t_{2g}} d^+_{a\alpha\sigma} d_{a\alpha\sigma} + \sum_{\alpha \in e_g} E_{e_g} d^+_{a\alpha\sigma} d_{a\alpha\sigma} \right) + H_U \tag{S2}$$

$$H_L = E_p \sum_b^{u,d} \sum_{\beta\sigma} p^+_{b\beta\sigma} p_{b\beta\sigma} \tag{S3}$$

$$H_t = \sum_a^{l,r} \sum_b^{u,d} \sum_{\alpha\beta\sigma} (t_{a\alpha b\beta\sigma} p^+_{b\beta\sigma} d_{a\alpha\sigma} + \text{h.c.}) \tag{S4}$$

$$H_{so} = \lambda_M \sum_a^{l,r} (\mathbf{S}_a \cdot \mathbf{L}_a) \tag{S5}$$

In addition to the crystal field splitting, the transition-metal d levels are split by the crystal field into $t_{2g}$ and $e_g$ orbitals with the energy difference $\Delta_{cf}$ between the energy levels $E_{t_{2g}}$ and $E_{e_g}$. The Hamiltonian $H_M$ for the transition metal also contains an effective Zeeman field, which originates from the local Coulomb repulsion and Hund's-rule coupling in the magnetically ordered phase:

$$H_U = -U \sum_{a=l,r} \sum_\sigma \mathbf{m}_a \cdot \mathbf{s}_{a\alpha} \tag{S6}$$

In the hopping term $H_t$, the hybridization matrix t depends on the d and p orbitals involved (corresponding to σ and π bonding of the orbitals) and also on their relative positions (left or right transition-metal M and up or down ligand L). $H_{so}$ describes the spin-orbit interaction within the magnetic d orbitals. The energy scheme is illustrated in Fig. S1(b). Unless otherwise stated, we will use the following reasonable parameters: $E_{t_{2g}} = 0$, $E_{e_g} = 2$ eV, $E_p = -5$ eV, $U = 10$ eV, $t_{pd\sigma} = -1.6$ eV, $t_{pd\pi} = 0.6$ eV, and $\lambda_M = 0.048$ eV.

We diagonalize the total Hamiltonian and then calculate the dipole moment using the occupied states. Our calculations show that, in terms of the local coordinate system (X′, Y′, Z′) defined in Fig. S1(a), the matrix **M** of the inter-site polarization is written as

$$\mathbf{M} = \begin{bmatrix} 0 & 0 & 0 \\ 0 & 0 & (\mathbf{P}_{12}^{X'Y'})_{Y'} \\ 0 & (\mathbf{P}_{12}^{Z'X'})_{Z'} & 0 \end{bmatrix}, \tag{S7a}$$

which is a consequence of the $D_{2h}$ symmetry of the four-atom cluster model. Our calculations show that $(\mathbf{P}_{12}^{X'Y'})_{Y'}$ is always much larger than $-(\mathbf{P}_{12}^{Z'X'})_{Z'}$ and is independent of the model parameters, in contrast to the case of the KNB model in which $(\mathbf{P}_{12}^{X'Y'})_{Y'} = -(\mathbf{P}_{12}^{Z'X'})_{Z'}$. To gain further insight into this finding, we examine how the d-states of $M_l$ and $M_r$ interact with the p-states of $L_u$ and $L_d$ [see Fig. S1(b)] in the absence and presence of SOC. It is found that the nonzero $(\mathbf{P}_{12}^{X'Y'})_{Y'}$ arises from the SOC-induced orbital mixing between the minority-spin $d_{z^2}$ orbital of $M_l$ and the $p_{x'}$ of $L_u$ [Fig. S1(c)], and the nonzero $(\mathbf{P}_{12}^{Z'X'})_{Z'}$ from that between the minority-spin $d_{x'^2-y'^2}$ of $M_l$ and the $p_{z'}$ of $L_u$. The extent of the orbital mixing can be described by the density matrix **D** with matrix elements $D_{mn}$ defined by $D_{mn} = \sum_i^{occ} \sum_{m,n} C_{im}^* C_{in}$, where $C_{im}$ is the coefficient of the local atomic basis $m$ in the $i$-th occupied state. We consider two cases in which the spins rotate in different planes. For the case when the two spins are in the X′Y′-plane (X′Z′-plane), the density matrix element between the $d_{z^2}$ orbital of $M_l$ and the $p_{x'}$ of $L_u$ (between the $d_{x'^2-y'^2}$ of $M_l$ and the $p_{z'}$ of $L_u$) is plotted as a function of the angle between the two spins in Fig. S1(d). As can be seen, both density matrix elements exhibit a sinusoidal dependence, and the density matrix element is much larger for the X′Y′-plane than for the X′Z′-plane case. The large difference between $(\mathbf{P}_{12}^{X'Y'})_{Y'}$ and $-(\mathbf{P}_{12}^{Z'X'})_{Z'}$ reflects the structural anisotropy associated with the planar four-atom cluster; the two metal ions are bridged by two ligands with the Y′ axis in the plane of the cluster, whereas the Z′ axis is out of the plane.

The form of the matrix **M** obtained from density functional calculations can now be understood. In the local coordinate system (X, Y, Z) defined in the lower-right inset of Fig. 1(a), which results from the anticlockwise rotation of the local coordinate system (x, y, z) around the x axis by 137°, the matrix **M** of Eq. 5 determined from density functional calculations is rewritten as

$$\mathbf{M} = \begin{bmatrix} -4.8 & 0 & 0 \\ 0 & 6.8 & 79.6 \\ 0 & -13.9 & 6.8 \end{bmatrix}. \quad\quad (S7b)$$

in units of $10^{-5}$ eÅ. The Y axis is close to the distance vector between two edge-shared I atoms. In the local (X, Y, Z) coordinate system, $(\mathbf{P}_{12}^{XY})_{Y} = 79.6\times10^{-5}$ eÅ is much larger than $-(\mathbf{P}_{12}^{ZX})_{Z} = 13.9\times10^{-5}$ eÅ. In tight-binding calculations using the (X′, Y′, Z′) coordinate system, $(\mathbf{P}_{12}^{X'Y'})_{Y'}$ is much larger than $-(\mathbf{P}_{12}^{Z'X'})_{Z'}$, as discussed above. The (X, Y, Z) coordinate system is close to the (X′, Y′, Z′) coordinate system. This explains why $(\mathbf{P}_{12}^{XY})_{Y}$ is much larger than $-(\mathbf{P}_{12}^{ZX})_{Z}$ from our density functional calculations.

Due to the D$_{2h}$ symmetry of the four-atom cluster Mn$_2$I$_2$, the nonzero elements of the matrix **M** are that $(\mathbf{P}_{12}^{X'Y'})_{Y'}$ and $(\mathbf{P}_{12}^{Z'X'})_{Z'}$. In Fig. S2, we show the dependence of the polarizations on the various tight-binding parameters. In examining the dependence of one parameter, all the other parameters are fixed. We find that both $(\mathbf{P}_{12}^{X'Y'})_{Y'}$ and $-(\mathbf{P}_{12}^{Z'X'})_{Z'}$ increase monotonously with the SOC strength $\lambda_M$, the hopping parameter t, 1/U, and the crystal field splitting $\Delta_{cf}$. In particular, the polarizations increase almost linearly with $\lambda_M$. These results can be easily understood because the increase in these parameters enhances the mixing between the unoccupied orbitals and occupied orbitals. An interesting finding from Fig. S1 is that $(\mathbf{P}_{12}^{X'Y'})_{Y'}$ is always much larger than $-(\mathbf{P}_{12}^{Z'X'})_{Z'}$. This is different from the KNB model in which $(\mathbf{P}_{12}^{X'Y'})_{Y'} = -(\mathbf{P}_{12}^{Z'X'})_{Z'}$.

Reference

1. C. Jia *et al.*, Phys. Rev. B 74, 224444 (2006); C. Jia *et al.*, Phys. Rev. B 76, 144424 (2007).

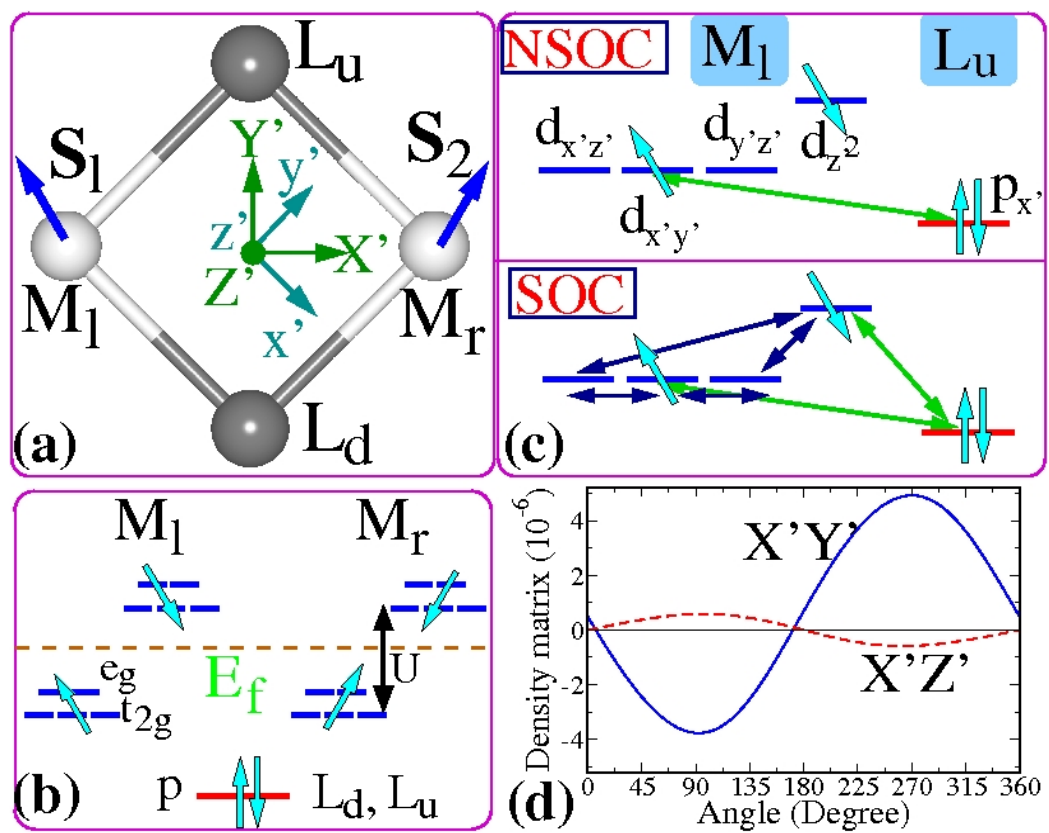

Figure S1. (Color online) (a) The $M_2L_2$ four-atom cluster model used for tight-binding analysis. The spin directions and the two coordinate systems used are indicated. (b) A schematic representation of the up-spin and down-spin d-states of $M_l$ and $M_r$ lying above the p-states of $L_u$ and $L_d$. (c) The interaction between the d-states of the metal $M_l$ with the p-states of the ligand $L_u$ in the absence of SOC (the upper panel labeled as NSOC) and in the presence of SOC (the lower panel labeled as SOC). The green double-headed arrows indicate the allowed interactions between $M_l$ and $L_u$, and the blue double-headed arrows indicate the allowed interactions among the d-states in the presence of SOC. (d) The density matrix between the $d_{z^2}$ orbital of $M_l$ and the $p_{x'}$ of $L_u$ for the case when these two spins are in the X′Y′-plane, and that between the $d_{x^2-y^2}$ orbital of $M_l$ and the $p_{z'}$ of $L_u$ for the case when these two spins are in the X′Z′-plane, as a function of the angle between the two spins. The left spin is fixed to be along the X-direction.

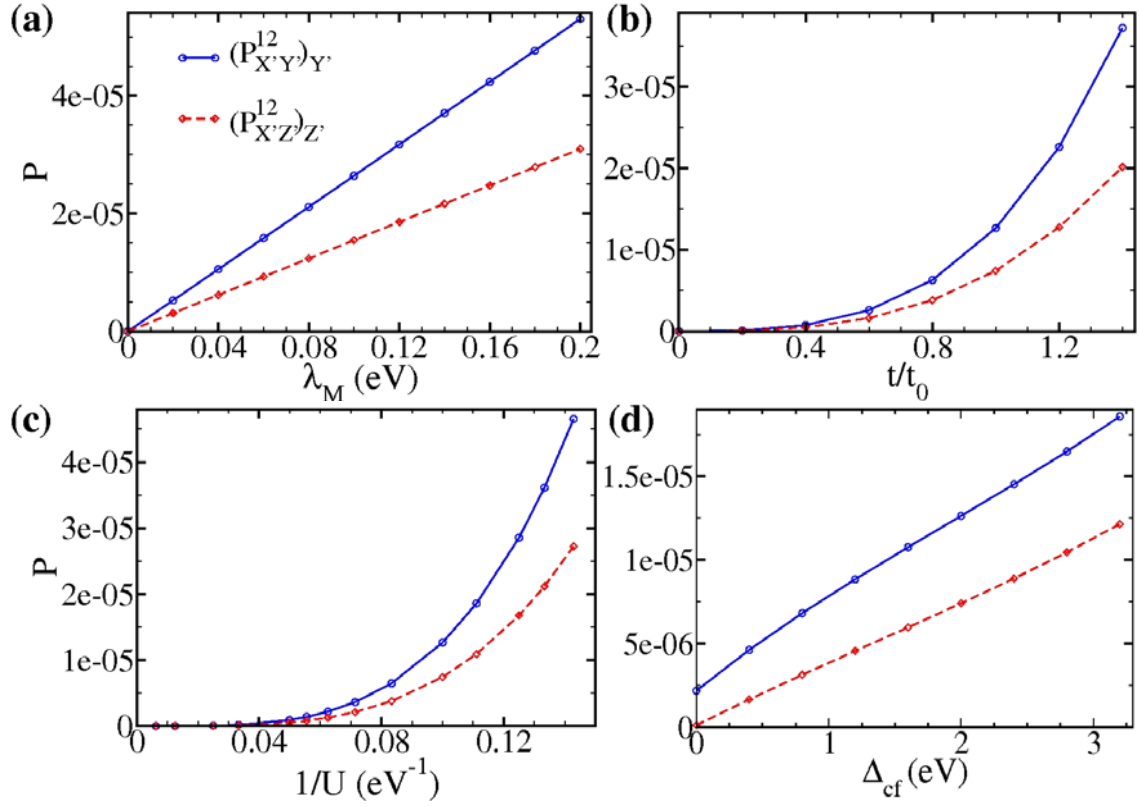

Figure S2. The dependence of the polarization $(\mathbf{P}_{12}^{X'Y'})_{Y'}$ and $(\mathbf{P}_{12}^{Z'X'})_{Z'}$ on (a) the SOC strength $\lambda_M$, (b) the scaled hopping parameter $t/t_0$, (c) the inverse of Hubbard U, and (d) the crystal field $\Delta_{cf}$.